\documentstyle[12pt]{article}
\def\bm#1{\mbox{\boldmath{$#1$}}}
\newcommand{\cftnote}{\renewcommand{\thefootnote}{\fnsymbol{footnote}}}
\newcommand{\resetftnote}{\setcounter{footnote}{0}}
\begin{document}

\begin{flushright}
UTTG-15-92\\
July 1992
\end{flushright}

\vspace{2mm}

\begin{center}
\cftnote
{\large \bf LAGRANGIANS FOR THE W-ALGEBRA MODELS}
\footnote{Research supported in part by the Robert A. Welch Foundation and NSF
Grant PHY 9009850}\\[7mm]
Jos\'e C. Gaite
\footnote{Fulbright Fellow.
Address after August, 1992: Departamento
de F\'{\i}sica, Universidad de Salamanca, Salamanca 37008, Spain.}
\\[4mm]
{\it Theory Group\\ Department of Physics\\ University of Texas\\
Austin, Texas 78712\\Bitnet: Gaite@utaphy}\\
\resetftnote
\end{center}

\vspace{2mm}

\begin{abstract}
The field algebra of the minimal models of W-algebras is amenable
to a very simple description as a polynomial algebra generated by few
elementary fields, corresponding to order parameters. Using this description,
the complete Landau-Ginzburg lagrangians for these models are obtained.
Perturbing these lagrangians we can explore their phase diagrams,
which correspond to multicritical points with $D_n$ symmetry.
In particular, it is shown that there is a perturbation for which the phase
structure coincides with that of the IRF models of Jimbo et al.
\end{abstract}

\section{Introduction}
The classification of two-dimensional conformal field theories
(2dCFT) is a flourishing branch of mathematical physics. It can
be applied in essentially two different directions. One is string
theory; the other is phase transitions in two dimensions and,
hopefully, also higher dimensions. With regards to this second
application, the information provided by 2dCFT may seem excessive.
In phase transition physics, before addressing the question of
critical exponents and, therefore, dimensions of fields, one
is usually more interested in determining the phase diagram.
For this purpose, the essential information consists of the
order parameters (with their symmetry) and the relevant fields
formed out of them. We are used to seeing this information coming from
the Landau potential, or its field-theory version, the Landau-Ginzburg
lagrangian. This lagrangian
is not a datum of 2dCFT but it can be obtained from it, as was
shown by A.B. Zamolodchikov \cite{ZamPot}
in the simple case of the unitary series of minimal models.
All these models have only one order parameter and $Z_2$
symmetry. In order to describe phase transitions with higher symmetry,
one is compelled to consider 2dCFT with an extended algebra.
Among them, the minimal models of the W-algebras \cite{FatLyk}
are the natural generalization of the Virasoro minimal models.
\par
The Zamolodchikov's procedure to associate lagrangians to
2dCFT begins by giving a polynomial structure to
the algebra of primary fields. He found that all the relevant
primary fields of a minimal model can be expressed as
composite (powers) of the
most relevant one (elementary field), which plays the role of order parameter.
This is a necessary condition for the existence of a
lagrangian and it is actually sufficient to guess it.
Nevertheless, there is a direct way to obtain it, relying on the
fact that the composite field next to the most relevant field
has to be identified with the descendant of the elementary
field, giving therefore the equation of motion.
In this paper, we will generalize this procedure to
the minimal models of the W-algebras.
An attempt in this direction was already made by I. Koh and S. Yang
\cite{KoYa},
but they did not obtain the complete structure of the
algebra of primary fields and their lagrangians missed
some important terms.

The Theory of Catastrophes is relevant in the context of Landau potentials, as
a mathematical framework in which many intuitive concepts of phase transitions
can be precisely formulated. Its classificatory power has already found wide
application in 2d $N=2$ superconformal theories. However, it is still a rather
specialized tool and we should not assume that the reader is familiar with it.
Therefore, we will put it aside, though making some remarks in footnotes.
\par
The paper is divided in three parts. We begin
with the simplest case, namely the $W_{(3)}$ algebra minimal models.
In the first part, we explicitly identify all the relevant fields as
powers of two elementary ones and therefore
endow the algebra with a polynomial structure.
The equations of motion and the lagrangian ensue from
the identification of further powers of the elementary fields.
In the second part, the phase structure that results from
these potentials is briefly analyzed.
The third part is devoted to the general W-algebra models.
To deal with them one has to take
into account some new features that already appear for $W_{(4)}$.
The first model of the $W_{(4)}$ series has central charge $c = 1$
and belongs to the Ashkin-Teller critical line \cite{Yang} .
For its importance we take it as an example to study those
new features. We finally discuss the phase structure of
the general models and their relation with the solvable
statistical models defined by the japanese group \cite{JiMi} (JMO models).

\section {Structure of the field algebra of the $W_{(3)}$ models}

The minimal models of the $W_{(3)}$ algebra are constructed in
analogy with the current Virasoro minimal models, but using
instead a two-component free field \cite{FatZam}.
The procedure relies on
the Dotsenko-Fateev screening charge method. There are four
screening charges; the primary fields are
therefore labelled by a $2 \times 2$ matrix
\[\Phi\!\!\left(\!\!\begin{array}{cc} n& m\\
n'& m' \end{array}\!\!\right).\]
Every unitary model is still determined
by one integer $p$ and noted $W_{(3)}^p$. This model contains
$p(p-1)^2 (p-2)/12$
independent primary fields and has $D_3$ symmetry.
The fusion rules are derived
from the neutrality condition including definite
numbers of screening charges,

\begin{equation}
{\bm\alpha_{(3)}} = {\bm\alpha_{(1)}} + {\bm\alpha_{(2)}}
+ N \alpha_+ {\bf e}_1
+ N' \alpha_+ {\bf e}_2 + M \alpha_- {\bf e}_1
+ M' \alpha_- {\bf e}_2,    \label{fusion}
\end{equation}
with
\begin{eqnarray}
{\bm\alpha_{(i)}} \equiv
{\bm\alpha}\!\!\left(\!\! \begin{array}{cc} n_{(i)}& m_{(i)}\\
n'_{(i)}& m'_{(i)} \end{array}\!\!\right)
&=& ([1-n_{(i)}]\alpha_+ + [1-m_{(i)}]\alpha_-) {\bm\omega_1} + \nonumber\\
& & ([1-n'_{(i)}]\alpha_+ + [1-m'_{(i)}]\alpha_-) {\bm\omega_2}.
\end{eqnarray}
Here ${\bf e}_i$ and ${\bm\omega_i}$ are the positive roots and
fundamental weights of $SU(3)$ ($A_2$ algebra), respectively.
Note that Eq. (\ref{fusion})
expresses the Clebsch-Gordan decomposition of the tensor product of
two $SU(3)\otimes SU(3)$ representations with highest weights
$-{\bm\alpha_{(1)}}$ and $-{\bm\alpha_{(2)}}$.
The solution for $n_{(3)}, n'_{(3)}, m_{(3)}, m'_{(3)}$ is
\begin{eqnarray}
n_{(3)} &=& n_{(1)} + n_{(2)} - 1 -2N + N' \nonumber\\
n'_{(3)} &=& n'_{(1)} + n'_{(2)} - 1 -2N' + N   \label{fusion-n}
\end{eqnarray}
and similar equations for $m_{(3)}, m'_{(3)}$.

\subsection{Elementary and composite fields}

The most relevant fields are
\[\sigma=\Phi\!\!\left(\!\! \begin{array}{cc} 2& 2\\
1& 1 \end{array}\!\! \right),~~
\bar{\sigma}=\Phi\!\!\left(\!\! \begin{array}{cc} 1& 1\\
2& 2 \end{array}\!\! \right),\]
which represent the $D_3$ spin density and its conjugate \cite{FatZam}.
They are the obvious candidates for elementary fields. It is
straightforward to prove that a certain number of the next most
relevant fields are obtained from them by the
Zamolodchikov's method with the fusion rule (\ref{fusion}).
Besides, they arrange themselves in a triangular
structure corresponding to the $SU(3)$ lattice of dominant weights.
Before proceeding further, let us recall Zamolodchikov's
field identifications in more detail.
The primary field content of the
Virasoro minimal model $M_p$ can be visualized on a
grid of dimension $p \times (p-1)$ symmetric with respect
to the center. The elementary field is placed
at (2, 2) and its powers are placed on the main diagonal, up to
$(p-1, p-1)$. The next power is at $(p, p-2)$ or equivalently
at (1, 2). This is the lowest end of the second diagonal,
which contains the remaining powers, up to the $2p-4$th,
corresponding to the least relevant field.

In the present case, the conformal grid is four-dimensional
and the equivalent of the "main diagonal" is a two-dimensional
triangular section of it. To be precise, this "main diagonal" exactly
corresponds to the $SU(3)$ lattice of dominant
weights of level $p-3$, as it is proved by the identifications
\begin{equation}
\sigma^k \bar{\sigma}^l = \Phi\!\!\left(\!\!\begin{array}{cc} k+1& k+1\\
 l+1& l+1 \end{array}\!\! \right),
\end{equation}
with $0\leq k+l \leq p-3$, obtained from (\ref{fusion-n})
for the trivial $N=N'=M=M'=0$ case. The next power, $k+l=p-2$,
leads to essentially three different choices. For $l=0$,
$k=p-2$, the appropriate choice is $N=1, N'=M=M'=0$, and
\begin{equation}
\sigma^{p-2} = \Phi\!\!\left(\!\!\begin{array}{cc} p-3& p-1\\
 2& 1 \end{array}\!\! \right) =
\Phi\!\!\left(\!\!\begin{array}{cc} 2& 1\\
 1& 1 \end{array}\!\! \right).
\end{equation}
This choice holds for the subsequent powers filling
the triangle subtended from it up to the $2p-6$th power,
giving

\begin{equation}
\sigma^{p-2+k} \bar{\sigma}^l = \Phi\!\!\left(\!\!\begin{array}{cc} k+2& k+1\\
 l+1& l+1 \end{array}\!\! \right),
\end{equation}
with $0\leq k+l \leq p-4$. There are similar identifications
for $\bar{\sigma}^{p-2}$ and the triangle conjugate to the one
mentioned above, with $N=0,N'=1, M=M'=0$. We are left with
the middle triangle, which requires $N=1,N'=1, M=M'=0$. A convenient
form of the ensuing identifications is

\begin{equation}
\sigma^{p-k-2} \bar{\sigma}^{k+l-1} =
\Phi\!\!\left(\!\!\begin{array}{cc} k& k+1\\
 l+1& l \end{array}\!\! \right),
\end{equation}
with $1\leq k,l$ and $k+l\leq p-2$. In particular, for $k=1, l=p-3$,

\begin{equation}
(\sigma \bar{\sigma})^{p-3} = \Phi\!\!\left(\!\!\begin{array}{cc} 1& 2\\
 p-2& p-3 \end{array}\!\! \right) =
\Phi\!\!\left(\!\!\begin{array}{cc} 1& 2\\
 1& 2 \end{array}\!\! \right),
\end{equation}
we get the least relevant field of the thermal subalgebra
(containing the fields with $n=n'$ and $m=m'$). This
subalgebra is actually generated by its most relevant field,
the energy density
\[\epsilon =\sigma \bar{\sigma} = \Phi\!\!\left(\!\!\begin{array}{cc} 2& 2\\
 2& 2 \end{array}\!\! \right).\]
Its succesive powers span the subalgebra up to $\epsilon^{p-3}$,
identified above. This field will be important in the sequel. Its conformal
dimension, as obtained with the general formulas \cite{FatZam},
is $(p-2)/(p+1)$.

The three triangles containing the powers of the elementary field
ranging from the $p-2$nd to the $2p-6$th form a two-dimensional
section of the conformal grid equivalent to the second diagonal
of the Virasoro case. The field identifications above are consistent
across the borders, although in a nontrivial way. The fields
on the lower side of the upper triangle are
\begin{equation}
\sigma^{p-2} \bar{\sigma}^{l} =
\Phi\!\!\left(\!\!\begin{array}{cc} 2& 1\\
 l+1& l+1 \end{array}\!\! \right).
\end{equation}
They must me compared to the products of sigma with the fields on
the upper side of the middle triangle
\begin{equation}
\sigma^{p-3} \bar{\sigma}^{l} =
\Phi\!\!\left(\!\!\begin{array}{cc} 1& 2 \\
 l+1& l \end{array}\!\! \right).
\end{equation}
There is exact coincidince if the fusion with $\sigma$ is performed according
to equations (\ref{fusion-n})
with $M=1, N=N'=M'=0$,
\begin{equation}
\sigma(\sigma^{p-3} \bar{\sigma}^{l}) =
\Phi\!\!\left(\!\!\begin{array}{cc} 2& 1\\
 l+1& l+1 \end{array}\!\! \right).
\end{equation}

We know that in the Virasoro case, the main and second
diagonals of the conformal grid contain
all the relevant fields of the model. We have here a similar
property ; namely, the big triangle
with all the powers up to the $2p-6t$h contains all the relevant
fields of the $W_3^p$ model, as can be checked with the dimension
formula.

\subsection{Field equations and Landau-Ginzburg lagrangians}

The fields corresponding to the $2p-6$th power are specially
important. The most relevant is $\epsilon^{p-3}$. Pursuing
the analogy with the Virasoro case, we expect that further
products with $\sigma$ or $\bar{\sigma}$ must include the
lowest dimension descendants, namely
$\partial\sigma$ or $\partial\bar{\sigma}$, and therefore
originate just field equations. However, in contrast with
the Virasoro case, other fields already
identified appear before these descendants
in the regular part of the operator product expansion (OPE);
they have to be kept in the field equations. This new
feature was already remarked for the $p=4$ case in \cite{I}.
The OPE for $\epsilon^{p-3}$ in particular, can be written as
\begin{eqnarray}
\sigma(z)\: \epsilon^{p-3}(0) &=&
\Phi\!\!\left(\!\!\begin{array}{cc} 2& 2\\
 1& 1 \end{array}\!\! \right) \Phi\!\!\left(\!\!\begin{array}{cc} 1& 2\\
 1& 2 \end{array}\!\! \right) =
\frac{1}{z^{\Delta_{\epsilon^{p-3}}}}\: \sigma + \cdots \nonumber\\
&+& \frac{1}{z^{\Delta_\sigma+\Delta_{\epsilon^{p-3}}-\Delta_{\phi}}}\:
{\phi}
+ \frac{1}{z^{\Delta_{\epsilon^{p-3}}-1}}\: \partial\sigma + \cdots,
\label{OPEdes}
\end{eqnarray}
where we have shown the most singular term and the initial
regular terms; numerical coefficients have been omitted. The field
$\phi$, yet to be determined, must
have been already identified as
a power of $\sigma$ and $\bar{\sigma}$, and needs to be such that the
corresponding
exponent of $z$ is positive and smaller than $1-\Delta_{\epsilon^{p-3}}$,
that is to say
\[\Delta_{\phi} < 1 + \Delta_{\sigma} = 1 + \frac{4}{3p(p+1)}.\]
We notice that the fusion rule Eq. \ref{fusion-n} with $M=1, N=N'=M'=0$,
gives
\begin{equation}
\sigma \epsilon^{p-3} =
\Phi\!\!\left(\!\!\begin{array}{cc} 2& 1\\
1& 3 \end{array}\!\! \right) = \Phi\!\!\left(\!\!\begin{array}{cc} p-3& p-3\\
2& 1 \end{array}\!\! \right) = \sigma^{p-4} \bar{\sigma}^{p-2}
\end{equation}
with conformal dimension
\begin{equation}
\Delta\!\!\left(\!\!\begin{array}{cc} 2& 1\\
 1& 3 \end{array}\!\! \right) =
\frac{3p(p-1)+4}{3p(p+1)}.
\end{equation}
This field therefore satisfies the condition and can be taken as
$\phi$.

With $\phi$ identified as above, the OPE (\ref{OPEdes}) yields
the following field equation (Henceforth, the symbol $\simeq$ will stand
for equality up to numerical coefficients)
\begin{equation}
\partial^2\sigma \simeq \sigma^{p-4} \bar{\sigma}^{p-2}
+ \sigma(\sigma \bar{\sigma})^{p-3}.             \label{feq}
\end{equation}
This equation leads to the lagrangian
\begin{eqnarray}
{\cal L} &\simeq& \partial\sigma \partial\bar{\sigma} +
\sigma^{p-4} \bar{\sigma}^{p-1}
+ (\sigma \bar{\sigma})^{p-2} + c.c. \nonumber\\
&=& \partial\sigma \partial\bar{\sigma} +
(\sigma \bar{\sigma})^{p-4} (\sigma^3 +\bar{\sigma}^3) +
(\sigma \bar{\sigma})^{p-2}.            \label{lagrangian}
\end{eqnarray}
The new composites appearing in it can be identified as
\begin{equation}
(\sigma \bar{\sigma})^{p-2} = \Phi\!\!\left(\!\!\begin{array}{cc} 2& 1\\
 2& 1 \end{array}\!\! \right),
\end{equation}
\begin{equation}
\sigma^{p-4} \bar{\sigma}^{p-1} = \Phi\!\!\left(\!\!\begin{array}{cc} 1& 2\\
 4& 2 \end{array}\!\! \right),
\end{equation}
which are of course irrelevant fields. The first one is the least
irrelevant thermal field $\epsilon^{p-2}$ of conformal dimension
$\Delta_{\epsilon^{p-2}}=1+(3/p)$. The other is slightly
more irrelevant, $\Delta=1+[3(p+2)/p(p+1)]$, and is essential
to endow the lagrangian with the necessary $D_3$ symmetry.
Fields with higher dimension must also appear
in the lagrangian, for there are more field equations
coming from the products of $\sigma$ ($\bar{\sigma}$) and
other fields with the $2p-6$th power. Those additional
fields have to be neutral, and correspond to
monomial terms with $2p-4$th and $2p-5$th powers,
\[(\sigma \bar{\sigma})^{p-5} (\sigma^6 +\bar{\sigma}^6),
(\sigma \bar{\sigma})^{p-7} (\sigma^9 +\bar{\sigma}^9),\]
etc. These terms are important for the potential to be
well defined\footnote{This fact is specially clear in the
language of Catastrophe Theory, where the absence of those
terms produces an indeterminate potential \cite{Catas}.}
but, since they represent more irrelevant fields, a
renormalization-group argument would tell us
that their effect becomes negligible
close to the multicritical point. This circumstance can be mimicked
by assigning them small coefficients. Therefore,
they are not expected to play
any relevant role in the phase structure of the models and will be ignored
henceforth.
Nevertheless, the term $(\sigma \bar{\sigma})^{p-4}
(\sigma^3 +\bar{\sigma}^3)$ in (\ref{lagrangian}) needs to be kept, not only
to enforce the $D_3$ symmetry but also to prevent the potential from being
degenerate. The degeneracy of the potential consisting of just the
$(\sigma \bar{\sigma})^{p-2}$ term (the one found in \cite{KoYa}) manifests
itself as the possibility of obtaining under some perturbations an infinite
number of minima (one or several circumferences). Summarizing,
the lagrangian (\ref{lagrangian}) is the correct starting point for the study
of the phase diagram in the next section.

We would like to point out a peculiarity of the lowest-$p$ models:
Counting their primary fields one realizes that there are not
enough to support the identifications above. For instance, $W_{(3)}^4$ and
$W_{(3)}^4$ have 6 and 20 primary fields, respectively, which cannot account
for all the fields mentioned above. The six fields of the former are all
relevant and identifiable with the composite fields up to second power of
the elementary ones.
However, we know that this model coincides with the nondiagonal modular
invariant $(A_4,D_4)$ of the Virasoro series \cite{Li2,KoYa},
which has additional primary fields for higher powers of the elementary fields,
namely,
\begin{eqnarray}
\Phi_{(12,13)} = \sigma^3,~~\Phi_{(13,12)} = \bar{\sigma}^3 \\
\Phi_{(13)} = \epsilon^2 = (\sigma\bar{\sigma})^2.
\end{eqnarray}
Therefore, these fields have to be W-secondaries. It is easy to find that
indeed
\begin{eqnarray}
\sigma^3 = W\!_{-1}\,\epsilon,~~\bar{\sigma}^3 = \overline{W}\!_{-1}\,\epsilon
\\
\epsilon^2 = \overline{W}\!_{-1}\,W\!_{-1}\,\epsilon.
\end{eqnarray}
In conclusion, some of the composite fields of the low-$p$ models turn out to
be W-secondaries with the necessary symmetry properties.

\section{Phase structure of the $W_{(3)}$ models}

The mean-field phase diagram is obtained analyzing the various
configurations of minima produced upon perturbation of
the potential part of the lagrangian by relevant fields.
The topology of the mean-field phase diagram is
identical to the actual phase diagram (this is the reason why
the Landau potential suffices to study the phase structure).
A complete representation of the phase diagrams is
however extremely complicated, as can be imagined from the
difficulties already encountered for the simplest case,
the three-state Potts model $W^4_{(3)}$, worked out in \cite{I}.
Nevertheless, one is mostly interested in certain properties
of the phase diagram, namely, the number of possible phases
(stable states) and their topological interrelation, which only demand a
limited knowledge of it. These properties are
conveniently expressed by the state diagram, which we
define as a set of points corresponding to the possible states
(either stable or unstable)
in a definite region of the phase diagram, and being linked whenever they can
merge under a suitable perturbation\footnote{Hence,
for phase diagrams derived from a potential the state diagram coincides
with the Dynkin diagram of a level curve, as defined in \cite{Norberto}
and used in this context in \cite{I}.
However, we prefer here the name state diagram since we
will be using Lie-algebra terms, among which Dynkin
diagram already has a very standard meaning, not
identical to that in \cite{Norberto}. In fact,
the state diagrams for the $W_{(n)}$ models are related with {\em weight}
diagrams of the underlying Lie algebra.}.
This diagram provides a direct connection
with statistical mechanics: A perturbed conformal model can be
interpreted as
the renormalization-group universality class
of lattice models with a state variable
living on that diagram. Now we recall that the $W^p_{(n)}$
are known to describe the critical behavior of the JMO
lattice models \cite{JiMi}
(see also \cite{Bilal}). These are
interaction round a face (IRF) models
with state variables defined on the dominant-weight lattice
of some Lie algebra, as a generalization of the
restricted solid on solid (RSOS) models of
Andrews, Baxter and Forrester, the
critical behavior of which is described by
the Virasoro minimal models. A motivation for the construction of these IRF
models actually arose from the
representation of the minimal W models as cosets $(A_n \oplus A_n, A_n)$ with
level $(l-1,1)$, in which the $A_n$ dominant weights
of level $l$ play a natural role (recall the role of (\ref{fusion}) as
the Clebsch-Gordan decomposition rule).

The relation between the state diagram of the $W^4_{(3)}$ model and
the diagram of dominant fundamental weights of $A_2$ is apparent in \cite{I}.
It is to be expected that the state diagram of the
$W^p_{(3)}$ model coincides with the $A_2$ dominant-weight diagram of
level $p-3$. Therefore, we would be interested in obtaining
that state diagram by perturbing the potential in a suitable way.
Finding this perturbation in the general case turns out to be
quite a nontrivial problem: Contrary to the Virasoro case, with
one order parameter, for which is rather easy to discern the
effect of each individual perturbation, in the case with
several order parameters it is almost impossible to predict
how the potential will deform under a particular perturbation.
However, there are some rules, as we shall see. First of all, let us introduce
real variables, by $\sigma=x+iy$ ($\bar{\sigma} = x-iy$), more suitable
for representation of the potential, and
simplified notation for the two independent $D_3$ invariants
\[I_0 = \sigma \bar{\sigma} = x^2+y^2,\]
\[I_1 = \sigma^3 +\bar{\sigma}^3 = x^3-3xy^2.\]
The potential takes the form
of a polynomial in $I_0$ and $I_1$,
\begin{equation}
V(x,y) = I_0^{p-2} + I_0^{p-4} I_1 + {\cal P}(I_0,I_1), \label{pot}
\end{equation}
where the symmetric perturbation ${\cal P}(I_0,I_1)$ has degree $2p-6$
as a polynomial in $x,y$.
If the potential contained only $I_0$, it would have full $O(2)$
symmetry and depend on just the radial variable $r=\sqrt{x^2+y^2}$.
Although this possibility is not admissible in the present context,
it is convenient to compare with the case with one order parameter.
Here the maximum unfolding deformation is achieved
for non-null values of all the coupling constants in the symmetric
perturbation. The modification introduced by $I_1$ amounts to
reduce the symmetry of this rotational invariant potential
to $D_3$. The structure is still the same along the three directions
in the $(x,y)$ plane for which $I_1=0$, and must be qualitative
similar all over. Therefore, one has to tune carefully all the
coupling constants to produce minima configurations
corresponding to the desired state diagram.
This problem, topological in essence, cannot be formulated
in any simple algebraic form.
The simplest models can be succesfully handled by computer programs
capable of representing algebraic surfaces, such as Mathematica.
In this way, one tests different values of coupling constants
to observe how the potential deforms. With some insight and
a good deal of luck, one may arrive to a configuration recognizable
as described by the appropriate $A_2$ dominant-weight diagram.
Needless to say, the number of coupling constants increases
rapidly with $p$ and beyond $p = 6$ this trial-and-error method
ceases to be feasible.

Fortunately, there are more powerful topological arguments that can
be directly applied to solve the inverse problem; namely,
finding from the state diagram
a potential with a minima configuration that fits it. Afterwards, this
potential can be compared with the original one.
In order to expose the method, let us introduce the
concept of level curves for a potential; in our case,
$V(x,y)=c$ defines an algebraic curve at level\footnote
{Throughout the paper, we use the term level with two different meanings,
namely,
as the value of $z=V(x,y)$ in the $(x,y,z)$ coordinate representation
providing the graphical notion of level
curves or as the integer associated to a representation of a
Lie algebra.} $c$.
This curve is singular if there are extremal points of
the potential (minima, maxima or saddle points) at that level.
Now we consider the $A_2$ dominant-weight diagram of level $l$;
we mark the middle points on each link and
construct a singular curve joining these points, as in Fig. 1
\footnote{This process is the inverse of obtaining the
Dynkin diagram of a curve according to \cite{Norberto}.}.
We would like to identify this topological curve
with a level curve of the potential (\ref{pot})
for $l=p-3$. The first thing we observe is that the
curve is the product of a definite number of components
(two in Fig. 1). We further notice that there are precisely $(l+1)/2$
trefoil shaped curves when $l$ is odd and $l/2$ trefoil curves
plus one of circular type when $l$ is even.
The level curve of the potential for the $W^4_{(3)}$ model,
\begin{equation}
V(x,y) = I_0^2 + I_1 + w I_0 = 0,  \label{3st}
\end{equation}
is the simplest algebraic curve with the trefoil shape\footnote
{This potential has been described in a previous paper \cite{I}.
That (\ref{3st}) represents a trefoil can be understood without
calculation when $w=0$: $I_1=0$ gives just three straight lines intersecting at
the origin and producing six sectors where $V(x,y)$ takes negative and
positive values alternatively. The effect of adding $I_0^2$ is raising
the potential for $r$ large enough, hence closing the three negative sectors.};
(observe that there are two different topological
types depending on the sign of $w$).
The simplest circular curve is, of course, the circle $I_0 - c = 0$.
Therefore, the simplest algebraic curve which represents the
whole topological curve is given by the product
\begin{equation}
\prod_{i=1}^{(l+1)/2} (I_0^2 + I_1 + w_i I_0) =
I_0^{l+1} + I_0^{l-1} I_1 + \cdots = 0,
\end{equation}
for $l$ odd, and by
\begin{equation}
\prod_{i=1}^{l/2} (I_0^2 + I_1 + w_i I_0)(I_0-c) =
I_0^{l+1} + I_0^{l-1} I_1 + \cdots = 0,
\end{equation}
for $l$ even. We see that both reproduce the potential (\ref{pot})
when $l=p-3$.

We should remark that the construction above yields
values of the coupling constants (in fact, a range)
for which the state diagram
is the desired $A_2$ dominant-weight diagram. However, there are certainly
other values that produce the same diagram: All the
level curves obtained by this method are singular and
correspond to very special potentials with the saddle points at
the same level; we can
easily imagine deformations of the potential that move some
saddle points off that level without altering its topology.
There can actually be a simpler set of values, with some of
them null. Moreover, those potentials are special in yet another sense:
We have to contemplate the presence of terms
with higher powers of $I_1$ in the non-perturbed potential, as
we noted at the end of the first section. These terms, though,
have small coefficients and will not change the topological type
of the potential. In any case, that construction
provides a privileged
starting point to describe the phase diagram and can be used to
probe the effect of other terms. We have done so
with Mathematica, drawing various contour plots
for the $p=5$ and $p=6$ models. A characteristic plot clearly exhibiting
the state diagram appears in Fig. 2.

\section{The general $W_{(n)}$ case}

The generalization of the Dotsenko-Fateev construction to $W_{(n)}$
demands a free massless scalar field with $n-1$ components.
Therefore, the number of screening charges is $2(n-1)$ and
the primary fields are labelled by an equal number of integers.
The unitary models $W_{(n)}^p$ contain
\mbox{$[p!(p-1)!]/[n!(n-1)!(p-n)!(p-n-1)!]$} spinless primary fields
and they have $D_n$ symmetry \cite{FatLyk}.
The neutrality condition to derive the fusion rules is
the straightforward generalization of that in (\ref{fusion})
\begin{equation}
{\bm\alpha(3)} = {\bm\alpha(1)} + {\bm\alpha(2)}
+ \alpha_+ \sum_{i=1}^{n-1} N_i{\bf e}_i
+ \alpha_- \sum_{i=1}^{n-1} M_i{\bf e}_i,     \label{fusion-gen}
\end{equation}
with
\begin{equation}
{\bm\alpha(j)} \equiv
{\bm\alpha}(\{n_i{(j)}\}\mid\{m_i{(j)}\}) =
\sum_{i=1}^{n-1}([1-n_i(j)]\alpha_+ + [1-m_i(j)]\alpha_-) {\bm\omega_i}.
\end{equation}
Here ${\bf e}_i$ and ${\bm\omega_i}$ are the positive roots and
fundamental weights of $SU(n)$ (algebra $A_{n-1}$, respectively.
The fusion rule (\ref{fusion-gen})
represents the Clebsch-Gordan decomposition for $SU(n)\otimes SU(n)$.
To obtain the numerical fusion rules one is to substitute for the
roots in terms of the weights,
\begin{eqnarray}
{\bf e}_1 &=& 2{\bm\omega_1} - {\bm\omega_2}, \nonumber\\
{\bf e}_i &=& 2{\bm\omega_i} - {\bm\omega_{i-1}} - {\bm\omega_{i+1}},
{}~~~~i=2,\ldots,n-2,       \nonumber\\
{\bf e}_{n-1} &=& 2{\bm\omega_{n-1}} - {\bm\omega_{n-2}},
\end{eqnarray}
and solve for $n_i(3)$ and $m_i(3)$,
\begin{eqnarray}
n_1(3) &=& n_1(1) + n_1(2) - 1 -2N_1 + N_2,   \nonumber\\
n_i(3) &=& n_i(1) + n_i(2) - 1 -2N_i + N_{i-1} + N_{i+1},
{}~~~~i=2,\ldots,n-2,   \nonumber\\
n_{n-1}(3) &=& n_{n-1}(1) + n_{n-1}(2) - 1 -2N_{n-1} + N_{n-2}
\label{fusion-n-gen}
\end{eqnarray}
(There are similar equations for $m_i$).

The elementary fields are the $n-1$ spin fields corresponding to the
fundamental representations of $SU(n)$,
\[\sigma_k =
\Phi(1,\ldots,1,\underbrace{2}_k,1,\ldots,1 \mid
1,\ldots,1,\underbrace{2}_k,1,\ldots,1).\]
They support the $D_n$ representations
\[\sigma_k \rightarrow e^{2\pi ki/n} \sigma_k\]
\[\sigma_k \rightarrow \bar\sigma_k;\]
namely, the most relevant fields, $\sigma_1$ and
$\bar\sigma_1 \equiv \sigma_{n-1}$, support the standard
representation and the others the remaining two-dimensional
representations; (if $n$ is even, the central field forms a one-dimensional
representation). The crucial fact is that
all the other relevant fields can be formed
as powers of those $n-1$ elementary ones and arranged on the $SU(n)$
lattice of dominant weights of level $2(p-n)$.
Showing this in the analogous way to the
$W_{(3)}$ case, demands thinking of a $2(n-1)$-dimensional
conformal grid and various $(n-1)$-dimensional sections.
We will restrict ourselves to $n=4$ for
simplicity, given that no essentially new properties arise
for higher $n$. In this case we have three elementary fields,
\[\sigma_1= \Phi(1,0,0\mid 1,0,0),\]
\[\sigma_2= \Phi(0,1,0\mid 0,1,0),\]
\[\bar{\sigma}_1 = \Phi(0,0,1\mid 0,0,1),\]
the second one being real. The lattice of dominant weights
is an isosceles pyramid (Fig. 3).
The ``main diagonal" contains fields up to level $p-4$,
which are identified in a straightforward way, from (\ref{fusion-n-gen})
when all $N_i = M_i =~0 \hspace{1.5mm}(i=1,2,3)$, with the
products of the elementary fields up to maximum total power $p-4$,
\begin{equation}
\sigma_1^k \sigma_2^l \bar{\sigma}_1^m =
\Phi(k+1,l+1,m+1\mid k+1,l+1,m+1),~~~~k+l+m \leq p-4.
\end{equation}
The identification of the next power, $k+l+m = p-3$, demands that
some of the $N_i$ be non-null. For $\sigma_1$ and $\bar{\sigma}_1$ we have
choices analogous to those
in the $W_{(3)}$ case:
\begin{itemize}
\item $N_1 = 1$ and the others null for
      \begin{equation}
\sigma_1^{p-3} =
\Phi(p-4,2,1\mid p-2,1,1) =
\Phi(2,1,1\mid 1,1,1).
     \end{equation}
\item The conjugate equation when $N_3 = 1$ and the others null.
\item The analogous middle triangle is also obtained by $N_1 = N_3 = 1$
(the remaining null).
\end{itemize}
However, the new component $\sigma_2$
introduces further possibilities, the simplest of which are
\begin{eqnarray}
\sigma_1^{p-4} \sigma_2 &=&
\Phi(p-3,1,1\mid p-3,1,1)\: \Phi(1,2,1\mid 1,2,1)  \nonumber\\
&=& \Phi(p-4,1,1\mid p-3,2,1) = \Phi(1,1,2\mid 2,1,1),
\end{eqnarray}
with $N_1 = N_2 = 1$ (remaining null), and its conjugate.
Another interesting one is
\begin{eqnarray}
\sigma_2^{p-3} &=&
\Phi(1,p-3,1\mid 1,p-3,1)\: \Phi(1,2,1\mid 1,2,1) \nonumber\\
&=& \Phi(1,p-4,1\mid 1,p-2,1) = \Phi(1,2,1\mid 1,1,1),
\end{eqnarray}
with $N_1 = N_3 = 1, N_2 = 2$. All these choices hold
for the respective domains of pyramidal shape that are subtended
from the position of these fields in the big pyramid up to
level $2p-8$ containing all the relevant fields.

An exhaustive description of the field identification domains
and their matching conditions is cumbersome but not necessary
to find the Landau-Ginzburg lagrangian: Like in the $W_{(3)}$
case, within the fields of highest level
only the most relevant is needed. This field also belongs to
the thermal subalgebra, now generated by two fields,
\begin{equation}
\epsilon_1 = \Phi(1,0,1\mid 1,0,1) =
\sigma_1 \bar{\sigma}_1
\end{equation}
and
\begin{equation}
\epsilon_2 = \Phi(0,2,0\mid 0,2,0) \simeq
\sigma_1 \bar{\sigma}_1 + \sigma_2^2
\end{equation}
(Recall that the symbol $\simeq$ means equal up to numerical coefficients).
The seeked field is
\begin{equation}
\epsilon_1^{p-4} =
(\sigma_1 \bar{\sigma}_1)^{p-4} =
\Phi(1,1,1\mid 2,1,2).
\end{equation}
Upon product with the elementary field $\sigma_1$, we obtain
the field equation
\begin{equation}
\partial^2\sigma_1 \simeq \sigma_1^{p-5} \bar{\sigma}_1^{p-4} \sigma_2
+ \sigma_1(\sigma_1 \bar{\sigma}_1)^{p-4}
\end{equation}
(Note the similarity with (\ref{feq}), although the monomial for the field
$\phi$ is now different, consistently with the present $D_4$
symmetry). Hence the lagrangian
\begin{equation}
{\cal L} \simeq \partial\sigma_1 \partial\bar{\sigma}_1 +
(\sigma_1\bar{\sigma}_1)^{p-5} (\sigma_1^2 + \bar{\sigma}_1^2) \sigma_2
+ (\sigma_1 \bar{\sigma}_1)^{p-3}.
\end{equation}
These are the crucial terms that determine the phase structure. However, like
in the $W_{(3)}$ case, other
terms corresponding to more irrelevant fields are also required for the
potential to be well defined. In addition, we will consider again a symmetric
perturbation, formed by lower degree terms corresponding to relevant fields.
To simplify the aspect of all these terms, it is convenient again
to regard the potential as a polynomial in the basic invariants.
The invariants for the standard representation of $D_4$ and their expression
in terms of real variables ($\sigma_1=x+iy$) are
\[I_0 = \sigma_1 \bar{\sigma}_1 = x^2+y^2,\]
\[I_1 = \sigma_1^4 +\bar{\sigma}_1^4 = 2(x^4+y^4-6x^2y^2).\]
Now the presence of $\sigma_2$, forming a one-dimensional $D_4$
representation\footnote{The complete representation of $D_4$ for the $W_{(4)}$
models is the
direct sum $A_2 \oplus E$, in the notation of Landau-Lifshitz's textbooks.},
produces new invariants
\[I_2 = \sigma_2^2 = z^2~~~~~(z \equiv \sigma_2)\]
\[I_{12} = (\sigma_1^2 +\bar{\sigma}_1^2) \sigma_2 = 2(x^2-y^2)z\]
That is to say, four basic invariants altogether. However, they are not
independent, for they satisfy the relation
\begin{equation}
I_{12}^2 = (I_1+2I_0) I_2.
\end{equation}
This fact introduces some complications\footnote{For a short account on
Invariant Theory bearing on this problem, see \cite{Stanley}.} but it will
suffice here to say that it amounts to the possibility of substituting for any
power of $I_{12}$ (higher than the first), which in consequence will not appear
in the potential.
Therefore, the potential takes the form
\begin{equation}
V(x,y,z) = I_0^{p-3} + I_0^{p-5} I_{12} + {\cal P}(I_0,I_1,I_2,I_{12}),
\label{pot-4}
\end{equation}
where the symmetric perturbation
\[{\cal P}(I_0,I_1,I_2,I_{12}) = {\cal P}_0(I_0, I_1,I_2) +
I_{12}{\cal P}_1(I_0, I_1,I_2)\]
has degree $2p-8$ as a polynomial in $x,y,z$.
The irrelevant additional terms are $I_0^{p-5} I_1, I_0^{p-4} I_2, I_0^{p-5}
I_2^2, I_0^{p-7} I_{12} I_1, I_0^{p-6} I_1 I_2$, etc.., and will be ignored
henceforth.

\subsection{Phase structure and its relation with IRF models}

In the study of phase structure we would like to generalize the method
applied to the $W_{(3)}$
models. If it applies to the $W_{(4)}$ ones as well,
we should expect to have configurations of minima that
reproduce the succesive sets of level-$l$ dominant integral weights of $A_3$.
The simplest model, $W_{(4)}^5$, is again very illustrative.
We will take advantage of the fact that there is an alternative description of
this model: It is a special
point on the line of 2dCFT defined on the orbifolds of a circle with variable
radius. These 2dCFT, with $c=1$ and $D_4$ symmetry, correspond to critical
Ashkin-Teller models with variable 4-spin coupling
(\cite{Yang,Ginsparg}). They have two twist fields with conformal dimension
$(1/16,1/16)$, realizing the standard $D_4$ representation (our $x,y$), and one
field with dimension depending on the coupling constant $K$
as $1/8{\pi}K$ \cite{Yang}, which realizes the non-trivial one-dimensional
representation (in our model this field is $z$, with $\Delta_z = 1/12$).

Specifying in (\ref{pot-4}) and after performing a ${\pi}/4$ rotation on the
$xy$-plane to write the cubic invariant in a more suitable form,
the perturbed potential for the $W_{(4)}^5$ model can be written as
\begin{equation}
V(x,y,z) = I_0^{2} + I_{12} + wI_0 + w'I_2 = (x^2+y^2)^2 + xyz +
w(x^2+y^2) + w'z^2,
\label{4st}
\end{equation}
This potential is analogous to (\ref{3st}) except for the presence of two
independent energy perturbations; the cubic term determines likewise
its symmetry. The equation $I_{12}=0$ represents the three coordinate
planes dividing the space in eight sectors for which the potential takes minus
and plus signs alternatively. Now the effect of adding $I_0^2$ does not suffice
to close a surface and produce minima, for there is no term $z^4$, and the
potential remains unbounded
in this direction\footnote{Note that the algebraic surface
$(x^2+y^2+z^2)^2 + xyz = 0$
has a four-lobed tetrahedral shape and therefore posseses higher symmetry,
namely, the full group of the tetrahedron $T_d$ (Landau-Lifshitz notation),
isomorphic to the symmetric group $S_4$.}.
Nevertheless, the quadratic perturbation $z^2$ bounds the potential, giving
minima in pyramidal configuration.

Let us attempt to calculate the position of the minima of the full potential
(\ref{4st}) from the equations
\begin{eqnarray}
\frac{\partial V}{\partial x} &=& 4x(x^2+y^2)+yz+2wx = 0, \\
\frac{\partial V}{\partial y} &=& 4y(x^2+y^2)+xz+2wy = 0, \\
\frac{\partial V}{\partial z} &=& xy+2w'z = 0.
\end{eqnarray}
The last equation is particularly simple to solve, giving
\begin{equation}
z=-xy/2w',  \label{z}
\end{equation}
which substituted back into the others yields the same equations as those
coming from the two-variable potential with $D_4$ symmetry
\begin{eqnarray}
V(x,y) &=& (x^2+y^2)^2 + (2w')^{-1}x^2y^2 + w(x^2+y^2) \nonumber\\
&=& x^4+y^4+(2-\frac{1}{2w'})x^2y^2 + w(x^2+y^2).
\label{4st-2d}
\end{eqnarray}
We know that its extrema
are placed on the vertices of a square. For the full 3-variable potential,
the respective $z$ obtained from \ref{z} are identical in absolute value but
negative or positive according to whether $x$ and $y$ have the
same sign or not. Therefore, they form the vertices of an isosceles pyramid.
However, it is not possible to check if the dimensions of this pyramid
coincide with those of
the $A_3$ dominant-weight lattice of level one (as drawn in Fig. 3).
In fact, this question is
nearly meaningless, given that there is no relation between $z$ units and $x$
or $y$ units.
Since the potential is at most quadratic in $z$, implying
that this variable cannot affect critical behavior\footnote
{This property is essential in Catastrophe theory, where a quadratic potential
is the simplest Morse function, which can be added to any catastrophe to give
an equivalent one. The particular equivalence to which we are led in our case
discarding $z$
is $T_{244} \simeq X_9$ of Arnold's classification \cite{Arnold}.},
a healthy point of view is to disregard it altogether and consider just
(\ref{4st-2d}) as the whole potential.
This potential has already been associated to the critical line of the
Ashkin-Teller model \cite{Li}. This line arises as a result of the existence
of a marginal field, identified with the term $x^2y^2$, in that model\footnote
{In Catastrophe Theory the potential (\ref{4st-2d}) is called
double-cusp catastrophe ($X_9$)
and has aroused interest as the simplest catastrophe with a modal deformation
\cite{Arnold}, which is just the one caused by the $x^2y^2$ term. The physical
consequences of modality were studied in \cite{Schul}}.
Its coefficient $\alpha=2-(1/2w')$ is directly related to the coupling
constant ${\pi}K = (\alpha/2)+1$, that is to say to the dimension of the field
$z$. For $\alpha=0$ we have two decoupled Ising models. When $\alpha=2$
the symmetry augments to $O(2)$, representing the Kosterlitz-Thouless point of
the XY model. In principle, the parameter $\alpha$ can also take negative
values
as long as $\alpha > -2$. For instance, the value that gives the dimension of
the field $z$ in the $W_{(4)}^5$ model is $-2/3$.
However, at $\alpha=-1$ something singular happens: The
conformal dimension of $z$ decreases to 1/16, the same as that of $x$ or $y$,
and the symmetry augments to $S_4$, representing the 4-state Potts model
\cite{Yang}. The potential has to include now the $z^4$ term, becoming
\begin{equation}
V(x,y) = x^4+y^4+z^4+ xyz + w(x^2+y^2+z^2),
\label{4st-P}
\end{equation}
which was considered years ago for the 4-state Potts model in any dimension
\cite{Zia}.
For smaller values of $\alpha$, it is the term $x^4+y^4$ that has to be
dropped; hence, the symmetry reduces to $D_4$ again.

The reader may be wondering about the bearing of the previous arguments on the
models with $p>5$. Here the field $z$ is certainly not to be disregarded
but the symmetry considerations still stand. In any case, the
argument used for the $W_{(3)}$ potentials applies: If the
term with $I_{12}$ in (\ref{pot-4}) is omitted, the maximum unfolding consists
of minima on spherical surfaces and the perturbation that produces it is
easily found. The effect of $I_{12}$ is enforcing the symmetry to give the
pyramidal shape. A detailed proof would involve a
construction with algebraic surfaces on the selected $A_3$ diagram of
dominant weights. It is straightforward and will not be expounded here.
Finally, it is clear as well how to generalize these methods to higher values
of $n$.

\section{Conclusions}

There are two main utilities of the lagrangian approach to 2dCFT: First, it
endows the field algebra with further structure that constitutes a much simpler
picture of it. Second, it allows a direct study of the phase diagram that
does not need to deal with the complications inherent to other methods of
treating perturbed 2dCFT (Bethe ansatz, etc). It also provides with simple
objects,
such as the state diagram, that can be directly related to statistical models
known to have critical behavior described by those 2dCFT. It is therefore a
sort of link between statistical models and 2dCFT. In this context, it may be
of some help to people concerned with subjects that manifest themselves in both
areas, like quantum groups.

In this paper we have developed the lagrangian approach for the minimal models
of W-algebras, relying on the free field construction, which directly relates
them to Lie algebras (to be precise, Kac-Moody algebras). We have seen how all
the relevant fields of $W_{(n)}^p$ can be identified with monomials of the
elementary fields that fit on the $A_{n-1}$ dominant-weight
lattice of level $2(p-n)$. The main field equation follows from an OPE,
in the Zamolodchikov's way. A new feature, worth emphasizing, is that
one has to
save in that OPE two fields instead of one to get the correct field equation.
Furthermore, this seems to be the only requirement for consistency of the
lagrangian description. We have obtained afterwards the Landau-Ginzburg
lagrangians, noting that they posses the $D_n$ symmetry of these models.
The potential contains the necessary information to study the whole phase
diagram. We have however limited ourselves to the phase structure produced by
symmetric perturbations. In particular, we have shown that there exists a
perturbation that originates a state diagram identifyable with that defining
the statistical models of Jimbo et al. \cite{JiMi}, namely, the $A_{n-1}$
dominant-weight diagram of level $p-n$. On the other hand,
the phase transition
between definite regimes of these models has already been shown
to be described by the
W-models. In consequence, our result closes the loop, showing a two-way
equivalence.

Many possibilities remain to be explored. For instance, we have confined
ourselves to the W models that are diagonal modular invariants. Extending the
methods in this paper to the non-diagonal ones seems quite straightforward.
It would be interesting to see whether the resulting potentials fit in a
classification (ADE or similar) agreeing with that known for the non-diagonal
modular invariants of W models.

\section{Acknowledgments}

I would like to thank again K. Li for questions raised in a conversation
some time ago that triggered part of this research.

\newpage

{\large \bf Figure Captions}\\[10mm]

\begin{enumerate}
\item Level curve for a minima comfiguration with the shape of the $A_2$
weight diagram of level 3.
\item Contour Plot of the potential
$V(x,y) = 0.49 I_1^2 + (I_0^2+I_1)(I_0-0.033)$,
corresponding to $W_{(3)}^5$.
\item $A_3$ Lattice of dominant weights of level 3 with some field
identifications.
\end{enumerate}

\end{document}